\documentclass
[prl,10pt,letterpaper,twocolumn,bibnotes,notitlepage,final,superscriptaddress,balancelastpage,showpacs]{revtex4}

\usepackage{amsmath}
\usepackage{amsfonts}
\usepackage{amssymb}
\usepackage{graphicx}

\usepackage{paralist}

\begin{document}

\title{Experimental Quantum Cryptography with Qutrits}

\author{Simon Gr\"oblacher}
\affiliation{Institut f\"ur Experimentalphysik, Universit\"at Wien,
Boltzmanngasse 5, A--1090 Wien, Austria}

\author{Thomas Jennewein}
\affiliation{Institut f\"ur Quantenoptik und Quanteninformation
(IQOQI), \"Osterreichische Akademie der Wissenschaften,
Boltzmanngasse 3, A--1090 Wien, Austria}

\author{Alipasha Vaziri}
\affiliation{Physics Department, University of Maryland, College
Park, MD 20742, USA}

\author{Gregor Weihs}
\affiliation{Institute for Quantum Computing \& Department of
Physics, University of Waterloo 200, University Ave. W, Waterloo, ON
N2L 3G1, Canada}

\author{Anton Zeilinger}
\affiliation{Institut f\"ur Experimentalphysik, Universit\"at Wien,
Boltzmanngasse 5, A--1090 Wien, Austria}
\affiliation{Institut f\"ur
Quantenoptik und Quanteninformation (IQOQI), \"Osterreichische
Akademie der Wissenschaften, Boltzmanngasse 3, A--1090 Wien,
Austria}

\date{\today}

\begin{abstract}
We produce two identical keys using, for the first time, entangled
trinary quantum systems (qutrits) for quantum key distribution. The
advantage of qutrits over the normally used binary quantum systems
is an increased coding density and a higher security margin. The
qutrits are encoded into the orbital angular momentum of photons,
namely Laguerre-Gaussian modes with azimuthal index $l$ $+1$, $0$
and $-1$, respectively. The orbital angular momentum is controlled
with phase holograms. In an Ekert-type protocol the violation of a
three-dimensional Bell inequality verifies the security of the
generated keys. A key is obtained with a qutrit error rate of
approximately $10\%$.
\end{abstract}


\maketitle

The wish to protect information from unauthorized listeners has
driven humans from early mankind on to invent all sorts of
cryptographic schemes and encryption algorithms. The modern computer
age has made the security need as important and the difficulty of
breaking classical algorithm-based cryptography as easy as never. In
the last decades of the 20th century cryptography schemes were
proposed where the security relies on the laws of quantum
mechanics~\cite{Wiesner83,Bennett84,Bennett85,Ekert91}. An intruder
trying to listen in will always be detected. Because these schemes
establish identical secret keys in two remote locations they have
since become known under the term Quantum Key Distribution (QKD).
QKD has been experimentally performed using all sorts of systems,
applying various protocols, over distances of up to
$120$~km~\cite{Gobby04a,Takesue05,Stucki02,Gobby04b}. These
experiments are performed in the lab as well as in real-life
environments, such as the nightly sky of a metropolitan
city~\cite{Kurtsiefer02,Resch05,Peng05}. Even a secure bank transfer
has been performed~\cite{Poppe04} and commercial prototype systems
are already available, which underlines the need and usefulness of
QKD systems.

All experiments performed so far were based on two-dimensional
quantum systems (qubits). Only in recent years noteworthy research
efforts have been put into higher-dimensional quantum systems
(qudits), in particular multi-dimensional entanglement. Especially
their application in tests of quantum nonlocality and quantum
information processing have attracted substantial
interest~\cite{Barreiro05,Cinelli05,Fitzi01,Molina05,Walborn06,Yang05}.
For quantum cryptography the usage of higher-dimensional systems
offers advantages such as an increased level of tolerance to noise
at a given level of security and a higher flux of information
compared to the qubit cryptography schemes. In general a QKD
protocol is considered secure as long as the mutual information of
the two parties $A$ and $B$ exchanging the key is greater than the
mutual information of $A$ and $E$ (or $B$ and $E$), where $E$ is an
eavesdropper. The possible mutual information of an eavesdropper
with one of the observers is strictly related to the noise rate of
the protocol and therefore an upper noise bound for a secure key
distribution can be found. For the BB84 and the Ekert qubit schemes
the limit on the noise ratio is $14.6\,\%$~\cite{Fuchs97}, which may
be slightly improved with alternative qubit schemes, e.g. using the
full set of $d+1$ mutually unbiased bases (MUBs)~\cite{Brusz98}. In
contrast, for three-dimensional quantum systems (qutrits) the noise
may be as high as $22.5\,\%$~\cite{Durt03} for the Ekert based
protocol. Furthermore, because a larger alphabet is used, each
system contains more information than a two-dimensional
one~\footnote{This can easily be seen, as for binary systems one
needs 8 bits (1 byte) to encode the standard ASCII characters,
whereas using trinary systems 5.048 trits are sufficient.}. Here we
present QKD using entangled qutrits in an extended
Ekert scheme~\cite{Ekert91}, similar to the first QKD experiment
with entangled qubits, performed by Jennewein \textit{et
al.}~\cite{Jennewein00}. The security of the keys obtained is thereby
confirmed by violating a three-dimensional Bell-type inequality.

In the present work, the qutrits are encoded into the orbital
angular momentum (OAM) of photons in the Laguerre-Gaussian modes
$LG_{p,l}$, which are the solution of the paraxial wave equation in
its cylindrical coordinate representation. The index $p$ represents
the number of radial nodes and the index $l$ is the winding number,
with $2\pi l$ describing the change in phase on a closed path around
the propagation axis. Thus a mode with $p=l=0$ is a Gaussian mode.
Throughout this paper we only consider photons with $p=0$, which
span an infinite-dimensional Hilbert space.

It has been experimentally shown~\cite{Mair01,Caetano02}, and was
later theoretically confirmed~\cite{Franke-Arnold02,Torres05}, that
in the process of parametric down-conversion the orbital angular
momentum is conserved for each individual pump photon if all beams
are collinear. Moreover, it has been demonstrated~\cite{Mair01} that
the down converted photons are in an entangled state with respect to
the OAM, which can be transformed into the maximally entangled state
using local filtering~\cite{Vaziri03}. Therefore, using a pump beam
with a Gaussian profile one can obtain the maximally entangled state
\begin{equation}
\psi=\alpha|0\rangle |0\rangle+\beta|1\rangle
|2\rangle+\gamma|2\rangle |1\rangle, \label{State}
\end{equation}
with $\alpha=\beta=\gamma=\frac{1}{\sqrt{3}}$. Here $|1\rangle$ is
the $LG_{0,1}$ mode, $|2\rangle$ the $LG_{0,-1}$ mode and
$|0\rangle$ the Gaussian mode $LG_{0,0}$. Such a maximally entangled
state can violate a three-dimensional Bell-type
inequality~\cite{Collins02,Kaszlikowski02} and therefore local
realism:
\begin{align}
S_{3}=&P(A_{1}=B_{1})+P(A_{2}=B_{1}-1)+P(A_{2}=B_{2})+\nonumber\\
+&P(A_{1}=B_{2})-P(A_{1}=B_{1}-1)-P(A_{2}=B_{1})-\nonumber\\
-&P(A_{2}=B_{2}-1)-P(A_{1}=B_{2}+1)\leq2 \label{Inequality},
\end{align}
with
\begin{equation}
P(A_{a}=B_{b}+k)=\sum_{j=1}^{3}P(A_{a}=j,B_{b}=(j+k)\,\textrm{mod}
\,3)
\end{equation}
being the probabilities that the outcomes of observers $A$ and $B$
measuring $A_{a}$ and $B_{b}$ differ by $k$ (modulo $3$). The
observables $A_{1},A_{2}$ and $B_{1},B_{2}$ correspond to different
local analyzer settings. Note, that the local realistic bound for
inequality (\ref{Inequality}) is the same as for the standard CHSH
inequality~\cite{CHSH69}. The maximal violation for the maximally
entangled state is
$S_{3}^{\textrm{max}}=4/(6\sqrt{3}-9)\approx2.873$. It is
interesting to note that for certain non-maximally entangled states
quantum mechanics predicts an even higher violation, i.e.
$S_{3}^{\textrm{non-max}}=1+\sqrt{11/3}\approx2.915$~\cite{Acin02}.
The violation of (\ref{Inequality}) has been experimentally shown by
Vaziri \textit{et al.}~\cite{Vaziri02a}.

To realize QKD based on an extended three-dimensional Ekert scheme,
the observers $A$ and $B$ randomly switch between three settings of
their transformation holograms. $A_{1},A_{2}$ ($B_{1},B_{2}$) are
the settings to maximally violate inequality (\ref{Inequality}) (and
therefore check the security of the protocol) and $A_{3}$, $B_{3}$
are settings leading to perfect correlations and therefore are used
for key production. $A$ and $B$ choose their settings independently
and at random and also record their photon detections independently.
After sufficiently many measurement runs $A$ and $B$ compare their
hologram settings. One-ninth of the produced data can be used
for the key, while $\frac{4}{9}$ of the data are for the violation
of the Bell inequality and the remaining $\frac{4}{9}$ have to be
discarded. After this basis reconciliation $B$ publicly announces
his data for the Bell inequality check, and $A$ computes the value
of $S_{3}$. In the case that $S_{3}>2$, the key is secure and an
eavesdropper will not have gained any useful information on the
key~\footnote{Of course $A$ and $B$ can still apply all standard QKD
procedures like privacy amplification, etc.}.

\begin{figure*}[t!]
\centering
\includegraphics*[width=17.2cm]{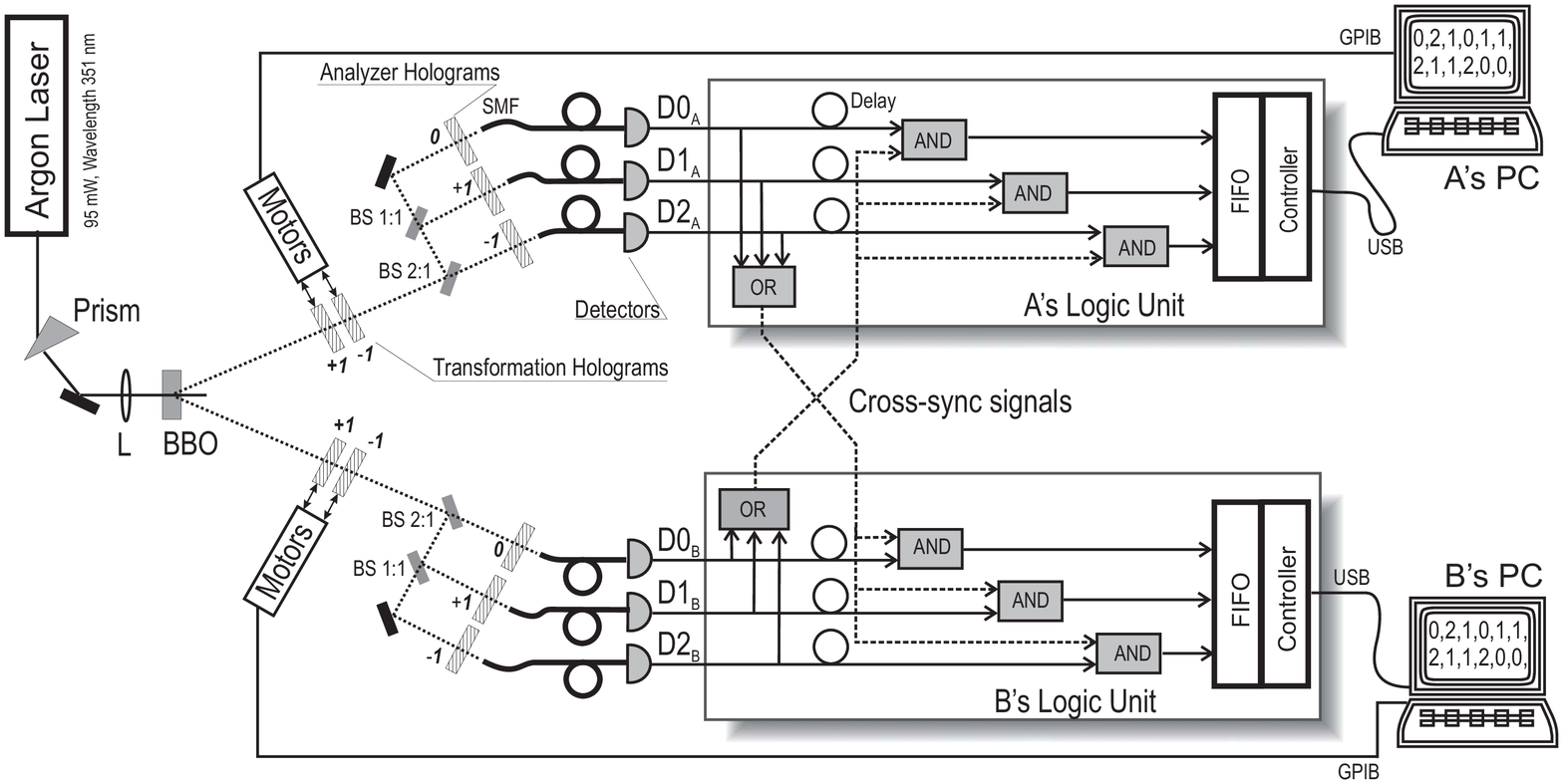}
\caption{\textit{Experimental Setup for the quantum key distribution
with qutrits. The source is an $\textrm{Ar}^{+}$ laser pumping a BBO
crystal at a wavelength of $351$~nm and an optical power of
approximately $95$~mW. Two phase holograms in each down-conversion
arm, mounted on computer controlled step motors, are used for
transforming the incoming maximally entangled qutrit state.
Probabilistic mode analyzers, consisting of beam splitters, mode
selection holograms and single mode fibers, allow the
differentiation between the three orthogonal modes $LG_{0,-1}$,
$LG_{0,0}$ and $LG_{0,1}$. The detection signals are then processed
in two separate logic units, where the coincidences are identified
via cross-sync signals. Depending on the local measurement result, a
value being either $0$, $1$ or $2$ is passed to the logics first-in
first-out buffer (FIFO) and read out by a computer.}} \label{Setup}
\end{figure*}

In our experimental setup (figure~\ref{Setup}) we pump a type-I $1.5$
mm thick $\beta$-barium-borate (BBO) crystal with an Ar$^{+}$ laser
at $351$ nm. The optical pump power is approximately $95$ mW and the
pump laser is vertically polarized. Via spontaneous parametric
down-conversion (SPDC) pairs of photons entangled in orbital angular
momentum are produced. To ensure indistinguishability, only the
energy-degenerate photons are selected via narrow band filters in
fibre-coupled air gaps. The produced state is almost maximally
entangled, with effective coefficients $\alpha=0.642\pm0.009$,
$\beta=0.546\pm0.009$ and $\gamma=0.539\pm0.009$, which are
calculated from the observed coincidence count rates for $A_{3}$ and
$B_{3}$.

In order to produce and control the $LG$ modes we use transmission
phase holograms
--- diffraction gratings which are interference patterns of an $LG_{0,1}$ mode with a
plane wave~\cite{Arlt98,Bazhenov90}. The holograms etched into
quartz glass are $3 mm\times3 mm$, have a periodicity of $30$ $\mu$m,
and their first-order diffraction efficiency at $702$ nm is
approximately $80\,\%$. If a beam passes such a hologram, an
$LG_{0,0}$ is, in the first diffraction order, transformed into an
$LG_{0,1}$ mode~\footnote{If the beam impinging on  the hologram is
an $LG_{0,1}$ mode it is likewise transformed into an $LG_{0,2}$
mode. Therefore, the holograms can be seen as approximate $+1$
ladder-operations for $LG$ modes.}. If the hologram is slightly
horizontally displaced, a superposition of the two modes is obtained,
with the respective amplitudes being a function of the
displacement~\cite{Mair01}. By inverting the beam direction the
transformation process of the hologram is also inverted and an
$LG_{0,1}$ is transformed into an $LG_{0,0}$. With these holograms
it is possible to create different superpositions of $LG$ modes
necessary for a test of Bell's inequality and for our cryptographic
scheme~\cite{Vaziri02b}.

To transform the state of the entangled photons, a pair of holograms
is placed in each down-conversion arm (see figure~\ref{Setup}). These
transformations approximate ladder-operations, i.e. one is a $+1$
and the other a $-1$ operation. Superpositions of the three $LG$
modes ($LG_{0,0}$, $LG_{0,1}$ and $LG_{0,-1}$) with different
relative amplitudes and phases can be produced by displacing the
individual holograms with step motors. Observers $A$ and $B$ now
choose the right positions of their holograms and can then violate
inequality~(\ref{Inequality}).

For the analysis of the different $LG$ modes the beams first pass a
$2:1$ and then a $1:1$ beam splitter, hence equally splitting them
into three parts. Each one of the resulting beams passes a hologram
and is then coupled into a single-mode fibre. Two of the holograms
are aligned such that they transform an $LG_{0,1}$ ($LG_{0,-1}$)
mode into an $LG_{0,0}$ mode. The third hologram is off-centred,
and therefore leaves the modes untransformed. Since only the
$LG_{0,0}$ has substantial overlap with the fibre mode this
arrangement acts as a probabilistic mode analyser with $\frac{1}{3}$
probability of success. The probabilistic nature of the analysers is
equivalent to a reduced detection efficiency but otherwise leads to
no additional security loopholes.

In order to find the optimal settings for the violation of the Bell
inequality each of the analyzer holograms was displaced by
$\pm1.2$~mm from the beam centre in $16$ equal steps. For every one
of the $83521$ ($17^{4}$) combinations of analyser settings all nine
coincidences and the single count rates were integrated over $5$~s
and written to a file. The data was finally analysed to check for
any violation of inequality~(\ref{Inequality}). The maximal value we
found for $S_{3}$ was $2.825\pm0.052$, which is a violation by
approximately $16$ standard deviations. The respective settings in
millimeters from the beam centre were $1.05$, $0.75$ (hologram $1$), $1.2$,
$0.3$ (hologram $2$) for $A$ and $0.45$, $1.05$ (hologram $3$),
$0.15$, $0.0$ (hologram $4$) for $B$'s side. The single count rates
were around $19000$ s$^{-1}$ and the coincidences of the perfect
correlations about $250$ s$^{-1}$, with a background of about
$7.4\,\%$, i.e. the sum over all coincidence counts in the unwanted
channels. In table~\ref{Table} some violations of~(\ref{Inequality})
and the corresponding hologram positions are shown.
\begin{table}
\begin{center}
\begin{tabular}{|c|c||c|c|c|c|}
\hline
$S_{3}$ & $\sigma$ & H$1$ [mm] & H$2$ [mm] & H$3$ [mm] & H$4$ [mm]\\
\hline \hline
$2.825$ & $0.052$ & $^{+1.05}_{+0.75}$ & $^{+1.2}_{+0.3}$ & $^{+0.45}_{+1.05}$ & $^{+0.15}_{\pm0.0}$\\
\hline $2.723$ & $0.052$ & $^{-0.15}_{-0.3}$ & $^{-0.3}_{\pm0.0}$ & $^{+0.45}_{+0.9}$ & $^{+0.15}_{-0.9}$\\
\hline $2.629$ & $0.056$ & $^{-0.15}_{-0.6}$ & $^{-0.6}_{-0.75}$ & $^{-0.15}_{-1.05}$ & $^{-0.6}_{-0.6}$\\
\hline
\end{tabular}
\end{center}
\caption{\textit{Experimental data for three exemplary Bell
parameters $S_{3}$, which violate the Bell inequality by several
standard deviations. The corresponding settings $A_{a}$ and $B_{b}$,
i.e. the horizontal displacements of the transformation holograms in
mm from the beam centre, are shown --- H$1$, H$2$ for $A$'s
holograms and H$3$, H$4$ for $B$'s.}} \label{Table}
\end{table}

The communication partners $A$ and $B$ had two different, completely
independent, computers and logics measuring their respective count
rates. They only identified coincidences with the help of
synchronisation signals. If they registered both, the signal from
the other side and a local detection, one entry, $0$, $1$ or $2$
depending on the result of the local detectors, was stored locally
in a computer file (see Fig.~\ref{Setup}). Furthermore, the current
setting of the transformation holograms was also written to the data
file. Each measurement lasted $1$~s and the step motors needed about
$5$~s to align. After many runs the data were analysed by comparing
the bases.

\begin{figure}[!]
\centering
\includegraphics*[width=8.6cm]{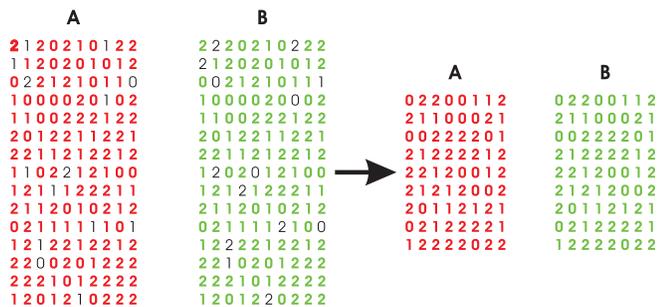}
\caption{\textit{On the left are the sifted keys
obtained by observers $A$ and $B$ via three-dimensional quantum key
distribution. The bold, coloured numbers are the correct trits while
the plain numbers are errors. From a total key of $150$ trits, $136$
entries ($90.7\,\%$) were the same for $A$ and $B$. The security of
this key is ascertained by the violation of the Bell
inequality~(\ref{Inequality}), with $S_{3}=2.688\pm0.171$. On the
right are the keys after a classical error reduction, which is done
by checking the parity of blocks of three trits and throwing away
those with different parities. The final keys are reduced to a
length of $72$ trits and are error-free.}} \label{Key}
\end{figure}

The Bell parameter was $S_{3}=2.688\pm0.171$, which represents a
clear violation of local realism. This ascertained the security of
the protocol. We extracted keys of a length of 150 trits for $A$ and
$B$ separately (the keys are shown in Fig.~\ref{Key}). Out of the
$150$ trits $14$ were errors, which corresponds to a quantum trit 
error rate (QTER) of $9.3\,\%$. This demonstrates the successful key
distribution, since Bell's inequality~(\ref{Inequality}) is violated
and additionally the error rate is well below the maximal allowed
noise ratio of $22.5\,\%$. Table~\ref{Table:Message} shows a
possible communication between $A$ and $B$ using the key generated
with the presented QKD.
\tabcolsep1pt
\begin{table*}[tbp]
\centering
\scriptsize{\begin{tabular}{||c|c c c c c c c c c c c c c c c c c c
c c c c c||} \hline \hline Original Text
&\,\textbf{T}&\textbf{H}&\textbf{E}&\textbf{\textvisiblespace}&\textbf{R}&\textbf{E}&\textbf{S}&\textbf{U}
&\textbf{L}&\textbf{T}&\textbf{\textvisiblespace}&\textbf{I}&\textbf{S}&\textbf{\textvisiblespace}&\textbf{F}&\textbf{O}&\textbf{R}&\textbf{T}
&\textbf{Y}&\textbf{\textvisiblespace}&\textbf{T}&\textbf{W}&\textbf{O}\\
Original Code &\,201&021&011&222&122&011&200&202&102&201&222&022&200&222&012&112&122&201&220&222&201&211&112\\
Key $A$ &\,022&001&122&110&002&100&222&201&212&222&122&212&001&221&212&002&201&121&210&212&222&122&222\\
\parbox[t]{2.3cm}{Cipher (Code+Key)mod3} &\,220&022&100&002&121&111&122&100&011&120&011&201&201&110&221&111&020&022&100&101&120&000&001\\
$E$'s Text
&\,\textbf{Y}&\textbf{I}&\textbf{J}&\textbf{C}&\textbf{Q}&\textbf{N}&\textbf{R}&\textbf{J}&\textbf{E}&\textbf{P}
&\textbf{E}&\textbf{T}&\textbf{T}&\textbf{M}&\textbf{Z}&\textbf{N}&\textbf{G}&\textbf{I}&\textbf{J}
&\textbf{K}&\textbf{P}&\textbf{A}&\textbf{B}\\
Cipher &\,220&022&100&002&121&111&122&100&011&120&011&201&201&110&221&111&020&022&100&101&120&000&001\\
Key $B$ &\,022&001&122&110&002&100&222&201&212&222&122&212&001&221&212&002&201&121&210&212&222&122&222\\
\parbox[t]{2.4cm}{Decrypted Code (Cipher-Key)mod3} &\,201&021&011&222&122&011&200&202&102&201&222&022&200&222&012&112&122&201&220&222&201&211&112\\
Decrypted Text
&\,\textbf{T}&\textbf{H}&\textbf{E}&\textbf{\textvisiblespace}&\textbf{R}&\textbf{E}&\textbf{S}&\textbf{U}
&\textbf{L}&\textbf{T}&\textbf{\textvisiblespace}&\textbf{I}&\textbf{S}&\textbf{\textvisiblespace}&\textbf{F}&\textbf{O}&\textbf{R}&\textbf{T}
&\textbf{Y}&\textbf{\textvisiblespace}&\textbf{T}&\textbf{W}&\textbf{O}\\
\hline \hline
\end{tabular}}
\caption{\textit{Encryption and decryption of a short message sent
between the two partners $A$ and $B$ using the error-corrected key
obtained via the three-dimensional quantum key distribution. Three
trits are sufficient to represent each letter of the alphabet plus
the space character. An eavesdropper trying to intercept the message
only gets random characters and hence cannot obtain any information
on the original text, whereas observer $B$ uses his key to decypher
the original message.}} \label{Table:Message}
\end{table*}

We have, for the first time, realised an experimental qutrit QKD.
 The completely independent parties $A$ and $B$
produce keys secured by the violation of a three-dimensional Bell
inequality by more than $4$ standard deviations. The sifted keys had
an error rate of approximately $10\,\%$. The effective key rate was
rather low due to the slow motorized base change. This could be
improved by implementing the base transformation with fast devices
such as a spatial light modulator or electro optical switches. In
addition, with a biased choice of the positions of the
transformation holograms, the key production rate could be further
increased. An additional challenge is the distortion-free
transmission of OAM-encoded photons over large distances. The
possibilities of free-space and fibre links are still under
investigation, since atmospheric turbulences and mode crosstalk in
fibres have to be overcome. Gibson \textit{et al.} already demonstrated a
free-space link of photons with OAM over a distance of $15$
m~\cite{Gibson04}. Alternatively, encoding higher dimensions into
other degrees of freedom of photons, such as time
bins~\cite{Thew04}, or as suggested by Chen \textit{et al.}~\cite{Chen05}
using the four-dimensional entangled states recently realized
by~\cite{Cinelli05,Yang05}, might also be considered, as they can be
transported in fibre or free-space over long distances. For
cryptography schemes based on single qutrits, similar to the BB84
scheme, transformations between MUBs are
required. With our holographic OAM scheme it is certainly possible
to do such transformations and a protocol of this type is currently
under investigation. In contrast to the polarization degree of
freedom, in principle there is no limitation on the dimension of the
two photon entanglement and therefore an extension of the qutrit to
a more general qudit case seems feasible.

We thank Jay Lawrence, Johannes Kofler and Martin St\"utz for
discussions and comments on the manuscript. This work has been
supported by the Austrian Science Fund (FWF) within project SFB F15,
and by the European Comission project RamboQ.


\begin{thebibliography}{38}
\expandafter\ifx\csname
natexlab\endcsname\relax\def\natexlab#1{#1}\fi
\expandafter\ifx\csname bibnamefont\endcsname\relax
  \def\bibnamefont#1{#1}\fi
\expandafter\ifx\csname bibfnamefont\endcsname\relax
  \def\bibfnamefont#1{#1}\fi
\expandafter\ifx\csname citenamefont\endcsname\relax
  \def\citenamefont#1{#1}\fi
\expandafter\ifx\csname url\endcsname\relax
  \def\url#1{\texttt{#1}}\fi
\expandafter\ifx\csname urlprefix\endcsname\relax\def\urlprefix{URL
}\fi \providecommand{\bibinfo}[2]{#2}
\providecommand{\eprint}[2][]{\url{#2}}

\bibitem[{\citenamefont{Wiesner}(1983)}]{Wiesner83}
\bibinfo{author}{\bibfnamefont{S.}~\bibnamefont{Wiesner}},
  \bibinfo{journal}{SIGACT\ News} \textbf{\bibinfo{volume}{15}},
  \bibinfo{pages}{78} (\bibinfo{year}{1983}).

\bibitem[{\citenamefont{Bennett and Brassard}(1984)}]{Bennett84}
\bibinfo{author}{\bibfnamefont{C.~H.} \bibnamefont{Bennett}} \bibnamefont{and}
  \bibinfo{author}{\bibfnamefont{G.}~\bibnamefont{Brassard}},
  \bibinfo{journal}{Proceedings of IEEE International Conference on Computers,
  Systems and Signal Processing, Bangalore, India} pp.
  \bibinfo{pages}{175--179} (\bibinfo{year}{1984}).

\bibitem[{\citenamefont{Bennett and Brassard}(1985)}]{Bennett85}
\bibinfo{author}{\bibfnamefont{C.~H.} \bibnamefont{Bennett}} \bibnamefont{and}
  \bibinfo{author}{\bibfnamefont{G.}~\bibnamefont{Brassard}},
  \bibinfo{journal}{IBM\ Tech.\ Discl.\ Bull.} \textbf{\bibinfo{volume}{28}},
  \bibinfo{pages}{3153} (\bibinfo{year}{1985}).

\bibitem[{\citenamefont{Ekert}(1991)}]{Ekert91}
\bibinfo{author}{\bibfnamefont{A.~K.} \bibnamefont{Ekert}},
  \bibinfo{journal}{Phys.\ Rev.\ Lett.} \textbf{\bibinfo{volume}{67}},
  \bibinfo{pages}{661} (\bibinfo{year}{1991}).

\bibitem[{\citenamefont{Gobby et~al.}(2004{\natexlab{a}})\citenamefont{Gobby,
  Yuan, and Shields}}]{Gobby04a}
\bibinfo{author}{\bibfnamefont{C.}~\bibnamefont{Gobby}},
  \bibinfo{author}{\bibfnamefont{Z.~L.} \bibnamefont{Yuan}}, \bibnamefont{and}
  \bibinfo{author}{\bibfnamefont{A.~J.} \bibnamefont{Shields}},
  \bibinfo{journal}{Appl.\ Phys.\ Lett.} \textbf{\bibinfo{volume}{84}},
  \bibinfo{pages}{3762} (\bibinfo{year}{2004}{\natexlab{a}}).

\bibitem[{\citenamefont{Takesue et~al.}(2005)\citenamefont{Takesue, Diamanti,
  Honjo, Langrock, Fejer, Inoue, and Yamamoto}}]{Takesue05}
\bibinfo{author}{\bibfnamefont{H.}~\bibnamefont{Takesue}},
  \bibinfo{author}{\bibfnamefont{E.}~\bibnamefont{Diamanti}},
  \bibinfo{author}{\bibfnamefont{T.}~\bibnamefont{Honjo}},
  \bibinfo{author}{\bibfnamefont{C.}~\bibnamefont{Langrock}},
  \bibinfo{author}{\bibfnamefont{M.~M.} \bibnamefont{Fejer}},
  \bibinfo{author}{\bibfnamefont{K.}~\bibnamefont{Inoue}}, \bibnamefont{and}
  \bibinfo{author}{\bibfnamefont{Y.}~\bibnamefont{Yamamoto}},
  \bibinfo{journal}{New J.\ Phys.} \textbf{\bibinfo{volume}{7}},
  \bibinfo{pages}{232} (\bibinfo{year}{2005}).

\bibitem[{\citenamefont{Stucki et~al.}(2002)\citenamefont{Stucki, Gisin,
  Guinnard, Ribordy, and Zbinden}}]{Stucki02}
\bibinfo{author}{\bibfnamefont{D.}~\bibnamefont{Stucki}},
  \bibinfo{author}{\bibfnamefont{N.}~\bibnamefont{Gisin}},
  \bibinfo{author}{\bibfnamefont{O.}~\bibnamefont{Guinnard}},
  \bibinfo{author}{\bibfnamefont{G.}~\bibnamefont{Ribordy}}, \bibnamefont{and}
  \bibinfo{author}{\bibfnamefont{H.}~\bibnamefont{Zbinden}},
  \bibinfo{journal}{New\ J.\ Phys.} \textbf{\bibinfo{volume}{4}},
  \bibinfo{pages}{41.1} (\bibinfo{year}{2002}).

\bibitem[{\citenamefont{Gobby et~al.}(2004{\natexlab{b}})\citenamefont{Gobby,
  Yuan, and Shields}}]{Gobby04b}
\bibinfo{author}{\bibfnamefont{C.}~\bibnamefont{Gobby}},
  \bibinfo{author}{\bibfnamefont{Z.~L.} \bibnamefont{Yuan}}, \bibnamefont{and}
  \bibinfo{author}{\bibfnamefont{A.~J.} \bibnamefont{Shields}},
  \bibinfo{journal}{Electron.\ Lett.} \textbf{\bibinfo{volume}{40}},
  \bibinfo{pages}{1603} (\bibinfo{year}{2004}{\natexlab{b}}).

\bibitem[{\citenamefont{Kurtsiefer et~al.}(2002)\citenamefont{Kurtsiefer,
  Zarda, Halder, Weinfurter, Gorman, Tapster, and Rarity}}]{Kurtsiefer02}
\bibinfo{author}{\bibfnamefont{C.}~\bibnamefont{Kurtsiefer}},
  \bibinfo{author}{\bibfnamefont{P.}~\bibnamefont{Zarda}},
  \bibinfo{author}{\bibfnamefont{M.}~\bibnamefont{Halder}},
  \bibinfo{author}{\bibfnamefont{H.}~\bibnamefont{Weinfurter}},
  \bibinfo{author}{\bibfnamefont{P.~M.} \bibnamefont{Gorman}},
  \bibinfo{author}{\bibfnamefont{P.~R.} \bibnamefont{Tapster}},
  \bibnamefont{and} \bibinfo{author}{\bibfnamefont{J.~G.}
  \bibnamefont{Rarity}}, \bibinfo{journal}{Nature}
  \textbf{\bibinfo{volume}{419}}, \bibinfo{pages}{450} (\bibinfo{year}{2002}).

\bibitem[{\citenamefont{Resch et~al.}(2005)\citenamefont{Resch, Lindenthal,
  Blauensteiner, B\"ohm, Fedrizzi, Kurtsiefer, Poppe, Schmitt-Manderbach,
  Taraba, Ursin et~al.}}]{Resch05}
\bibinfo{author}{\bibfnamefont{K.~J.} \bibnamefont{Resch}},
  \bibinfo{author}{\bibfnamefont{M.}~\bibnamefont{Lindenthal}},
  \bibinfo{author}{\bibfnamefont{B.}~\bibnamefont{Blauensteiner}},
  \bibinfo{author}{\bibfnamefont{H.~R.} \bibnamefont{B\"ohm}},
  \bibinfo{author}{\bibfnamefont{A.}~\bibnamefont{Fedrizzi}},
  \bibinfo{author}{\bibfnamefont{C.}~\bibnamefont{Kurtsiefer}},
  \bibinfo{author}{\bibfnamefont{A.}~\bibnamefont{Poppe}},
  \bibinfo{author}{\bibfnamefont{T.}~\bibnamefont{Schmitt-Manderbach}},
  \bibinfo{author}{\bibfnamefont{M.}~\bibnamefont{Taraba}},
  \bibinfo{author}{\bibfnamefont{R.}~\bibnamefont{Ursin}},
  \bibnamefont{et~al.}, \bibinfo{journal}{Optics\ Express}
  \textbf{\bibinfo{volume}{13}}, \bibinfo{pages}{202} (\bibinfo{year}{2005}).

\bibitem[{\citenamefont{Peng et~al.}(2005)\citenamefont{Peng, Yang, Bao, Zhang,
  Jin, Feng, Yang, Yang, Yin, Zhang et~al.}}]{Peng05}
\bibinfo{author}{\bibfnamefont{C.-Z.} \bibnamefont{Peng}},
  \bibinfo{author}{\bibfnamefont{T.}~\bibnamefont{Yang}},
  \bibinfo{author}{\bibfnamefont{X.-H.} \bibnamefont{Bao}},
  \bibinfo{author}{\bibfnamefont{J.}~\bibnamefont{Zhang}},
  \bibinfo{author}{\bibfnamefont{X.-M.} \bibnamefont{Jin}},
  \bibinfo{author}{\bibfnamefont{F.-Y.} \bibnamefont{Feng}},
  \bibinfo{author}{\bibfnamefont{B.}~\bibnamefont{Yang}},
  \bibinfo{author}{\bibfnamefont{J.}~\bibnamefont{Yang}},
  \bibinfo{author}{\bibfnamefont{J.}~\bibnamefont{Yin}},
  \bibinfo{author}{\bibfnamefont{Q.}~\bibnamefont{Zhang}},
  \bibnamefont{et~al.}, \bibinfo{journal}{Phys.\ Rev.\ Lett.}
  \textbf{\bibinfo{volume}{94}}, \bibinfo{pages}{150501}
  (\bibinfo{year}{2005}).

\bibitem[{\citenamefont{Poppe et~al.}(2004)\citenamefont{Poppe, Fedrizzi,
  Ursin, B\"ohm, Lor\"unser, Maurhardt, Peev, Suda, Kurtsiefer, Weinfurter
  et~al.}}]{Poppe04}
\bibinfo{author}{\bibfnamefont{A.}~\bibnamefont{Poppe}},
  \bibinfo{author}{\bibfnamefont{A.}~\bibnamefont{Fedrizzi}},
  \bibinfo{author}{\bibfnamefont{R.}~\bibnamefont{Ursin}},
  \bibinfo{author}{\bibfnamefont{H.~R.} \bibnamefont{B\"ohm}},
  \bibinfo{author}{\bibfnamefont{T.}~\bibnamefont{Lor\"unser}},
  \bibinfo{author}{\bibfnamefont{O.}~\bibnamefont{Maurhardt}},
  \bibinfo{author}{\bibfnamefont{M.}~\bibnamefont{Peev}},
  \bibinfo{author}{\bibfnamefont{M.}~\bibnamefont{Suda}},
  \bibinfo{author}{\bibfnamefont{C.}~\bibnamefont{Kurtsiefer}},
  \bibinfo{author}{\bibfnamefont{H.}~\bibnamefont{Weinfurter}},
  \bibnamefont{et~al.}, \bibinfo{journal}{Optics\ Express}
  \textbf{\bibinfo{volume}{12}}, \bibinfo{pages}{3865} (\bibinfo{year}{2004}).

\bibitem[{\citenamefont{Barreiro et~al.}(2005)\citenamefont{Barreiro, Langford,
  Peters, and Kwiat}}]{Barreiro05}
\bibinfo{author}{\bibfnamefont{J.~T.} \bibnamefont{Barreiro}},
  \bibinfo{author}{\bibfnamefont{N.~K.} \bibnamefont{Langford}},
  \bibinfo{author}{\bibfnamefont{N.~A.} \bibnamefont{Peters}},
  \bibnamefont{and} \bibinfo{author}{\bibfnamefont{P.~G.} \bibnamefont{Kwiat}},
  \bibinfo{journal}{Phys.\ Rev.\ Lett.} \textbf{\bibinfo{volume}{95}},
  \bibinfo{pages}{260501} (\bibinfo{year}{2005}).

\bibitem[{\citenamefont{Cinelli et~al.}(2005)\citenamefont{Cinelli, Barbieri,
  Perris, Mataloni, and Martini}}]{Cinelli05}
\bibinfo{author}{\bibfnamefont{C.}~\bibnamefont{Cinelli}},
  \bibinfo{author}{\bibfnamefont{M.}~\bibnamefont{Barbieri}},
  \bibinfo{author}{\bibfnamefont{R.}~\bibnamefont{Perris}},
  \bibinfo{author}{\bibfnamefont{P.}~\bibnamefont{Mataloni}}, \bibnamefont{and}
  \bibinfo{author}{\bibfnamefont{F.~D.} \bibnamefont{Martini}},
  \bibinfo{journal}{Phys.\ Rev.\ Lett.} \textbf{\bibinfo{volume}{95}},
  \bibinfo{pages}{240405} (\bibinfo{year}{2005}).

\bibitem[{\citenamefont{Fitzi et~al.}(2001)\citenamefont{Fitzi, Gisin, and
  Maurer}}]{Fitzi01}
\bibinfo{author}{\bibfnamefont{M.}~\bibnamefont{Fitzi}},
  \bibinfo{author}{\bibfnamefont{N.}~\bibnamefont{Gisin}}, \bibnamefont{and}
  \bibinfo{author}{\bibfnamefont{U.}~\bibnamefont{Maurer}},
  \bibinfo{journal}{Phys.\ Rev.\ Lett.} \textbf{\bibinfo{volume}{87}},
  \bibinfo{pages}{217901} (\bibinfo{year}{2001}).

\bibitem[{\citenamefont{Molina-Terriza
  et~al.}(2005)\citenamefont{Molina-Terriza, Vaziri, Ursin, and
  Zeilinger}}]{Molina05}
\bibinfo{author}{\bibfnamefont{G.}~\bibnamefont{Molina-Terriza}},
  \bibinfo{author}{\bibfnamefont{A.}~\bibnamefont{Vaziri}},
  \bibinfo{author}{\bibfnamefont{R.}~\bibnamefont{Ursin}}, \bibnamefont{and}
  \bibinfo{author}{\bibfnamefont{A.}~\bibnamefont{Zeilinger}},
  \bibinfo{journal}{Phys.\ Rev.\ Lett.} \textbf{\bibinfo{volume}{94}},
  \bibinfo{pages}{040501} (\bibinfo{year}{2005}).

\bibitem[{\citenamefont{Walborn et~al.}(2006)\citenamefont{Walborn, Lemelle,
  Almeida, and Ribeiro}}]{Walborn06}
\bibinfo{author}{\bibfnamefont{S.~P.} \bibnamefont{Walborn}},
  \bibinfo{author}{\bibfnamefont{D.~S.} \bibnamefont{Lemelle}},
  \bibinfo{author}{\bibfnamefont{M.~P.} \bibnamefont{Almeida}},
  \bibnamefont{and} \bibinfo{author}{\bibfnamefont{P.~H.~S.}
  \bibnamefont{Ribeiro}}, \bibinfo{journal}{Phys.\ Rev.\ Lett.}
  \textbf{\bibinfo{volume}{96}}, \bibinfo{pages}{090501}
  (\bibinfo{year}{2006}).

\bibitem[{\citenamefont{Yang et~al.}(2005)\citenamefont{Yang, Zhang, Zhang,
  Yin, Zhao, \.Zukowski, Chen, and Pan}}]{Yang05}
\bibinfo{author}{\bibfnamefont{T.}~\bibnamefont{Yang}},
  \bibinfo{author}{\bibfnamefont{Q.}~\bibnamefont{Zhang}},
  \bibinfo{author}{\bibfnamefont{J.}~\bibnamefont{Zhang}},
  \bibinfo{author}{\bibfnamefont{J.}~\bibnamefont{Yin}},
  \bibinfo{author}{\bibfnamefont{Z.}~\bibnamefont{Zhao}},
  \bibinfo{author}{\bibfnamefont{M.}~\bibnamefont{\.Zukowski}},
  \bibinfo{author}{\bibfnamefont{Z.-B.} \bibnamefont{Chen}}, \bibnamefont{and}
  \bibinfo{author}{\bibfnamefont{J.-W.} \bibnamefont{Pan}},
  \bibinfo{journal}{Phys.\ Rev.\ Lett.} \textbf{\bibinfo{volume}{95}},
  \bibinfo{pages}{240406} (\bibinfo{year}{2005}).

\bibitem[{\citenamefont{Fuchs et~al.}(1997)\citenamefont{Fuchs, Gisin,
  Griffiths, Niu, and Peres}}]{Fuchs97}
\bibinfo{author}{\bibfnamefont{C.~A.} \bibnamefont{Fuchs}},
  \bibinfo{author}{\bibfnamefont{N.}~\bibnamefont{Gisin}},
  \bibinfo{author}{\bibfnamefont{R.~B.} \bibnamefont{Griffiths}},
  \bibinfo{author}{\bibfnamefont{C.-S.} \bibnamefont{Niu}}, \bibnamefont{and}
  \bibinfo{author}{\bibfnamefont{A.}~\bibnamefont{Peres}},
  \bibinfo{journal}{Phys.\ Rev.\ A} \textbf{\bibinfo{volume}{56}},
  \bibinfo{pages}{1163} (\bibinfo{year}{1997}).

\bibitem[{\citenamefont{Bru\ss}(1998)}]{Brusz98}
\bibinfo{author}{\bibfnamefont{D.}~\bibnamefont{Bru\ss}},
  \bibinfo{journal}{Phys.\ Rev.\ Lett.} \textbf{\bibinfo{volume}{81}},
  \bibinfo{pages}{3018} (\bibinfo{year}{1998}).

\bibitem[{\citenamefont{Durt et~al.}(2003)\citenamefont{Durt, Cerf, Gisin, and
  \.Zukowski}}]{Durt03}
\bibinfo{author}{\bibfnamefont{T.}~\bibnamefont{Durt}},
  \bibinfo{author}{\bibfnamefont{N.~J.} \bibnamefont{Cerf}},
  \bibinfo{author}{\bibfnamefont{N.}~\bibnamefont{Gisin}}, \bibnamefont{and}
  \bibinfo{author}{\bibfnamefont{M.}~\bibnamefont{\.Zukowski}},
  \bibinfo{journal}{Phys.\ Rev.\ A} \textbf{\bibinfo{volume}{67}},
  \bibinfo{pages}{012311} (\bibinfo{year}{2003}).

\bibitem[{\citenamefont{Jennewein et~al.}(2000)\citenamefont{Jennewein, Simon,
  Weihs, Weinfurter, and Zeilinger}}]{Jennewein00}
\bibinfo{author}{\bibfnamefont{T.}~\bibnamefont{Jennewein}},
  \bibinfo{author}{\bibfnamefont{C.}~\bibnamefont{Simon}},
  \bibinfo{author}{\bibfnamefont{G.}~\bibnamefont{Weihs}},
  \bibinfo{author}{\bibfnamefont{H.}~\bibnamefont{Weinfurter}},
  \bibnamefont{and}
  \bibinfo{author}{\bibfnamefont{A.}~\bibnamefont{Zeilinger}},
  \bibinfo{journal}{Phys.\ Rev.\ Lett.} \textbf{\bibinfo{volume}{84}},
  \bibinfo{pages}{4729} (\bibinfo{year}{2000}).

\bibitem[{\citenamefont{Mair et~al.}(2001)\citenamefont{Mair, Vaziri, Weihs,
  and Zeilinger}}]{Mair01}
\bibinfo{author}{\bibfnamefont{A.}~\bibnamefont{Mair}},
  \bibinfo{author}{\bibfnamefont{A.}~\bibnamefont{Vaziri}},
  \bibinfo{author}{\bibfnamefont{G.}~\bibnamefont{Weihs}}, \bibnamefont{and}
  \bibinfo{author}{\bibfnamefont{A.}~\bibnamefont{Zeilinger}},
  \bibinfo{journal}{Nature} \textbf{\bibinfo{volume}{412}},
  \bibinfo{pages}{313} (\bibinfo{year}{2001}).

\bibitem[{\citenamefont{Caetano et~al.}(2002)\citenamefont{Caetano, Almeida,
  Ribeiro, Huguenin, dos Santos, and Khoury}}]{Caetano02}
\bibinfo{author}{\bibfnamefont{D.~P.} \bibnamefont{Caetano}},
  \bibinfo{author}{\bibfnamefont{M.~P.} \bibnamefont{Almeida}},
  \bibinfo{author}{\bibfnamefont{P.~H.~S.} \bibnamefont{Ribeiro}},
  \bibinfo{author}{\bibfnamefont{J.~A.~O.} \bibnamefont{Huguenin}},
  \bibinfo{author}{\bibfnamefont{B.~C.} \bibnamefont{dos Santos}},
  \bibnamefont{and} \bibinfo{author}{\bibfnamefont{A.~Z.}
  \bibnamefont{Khoury}}, \bibinfo{journal}{Phys.\ Rev.\ A}
  \textbf{\bibinfo{volume}{66}}, \bibinfo{pages}{041801(R)}
  (\bibinfo{year}{2002}).

\bibitem[{\citenamefont{Franke-Arnold et~al.}(2002)\citenamefont{Franke-Arnold,
  Barnett, Padgett, and Allen}}]{Franke-Arnold02}
\bibinfo{author}{\bibfnamefont{S.}~\bibnamefont{Franke-Arnold}},
  \bibinfo{author}{\bibfnamefont{S.~M.} \bibnamefont{Barnett}},
  \bibinfo{author}{\bibfnamefont{M.~J.} \bibnamefont{Padgett}},
  \bibnamefont{and} \bibinfo{author}{\bibfnamefont{L.}~\bibnamefont{Allen}},
  \bibinfo{journal}{Phys.\ Rev.\ A} \textbf{\bibinfo{volume}{65}},
  \bibinfo{pages}{033823} (\bibinfo{year}{2002}).

\bibitem[{\citenamefont{Torres et~al.}(2005)\citenamefont{Torres,
  Molina-Terriza, and Torner}}]{Torres05}
\bibinfo{author}{\bibfnamefont{J.~P.} \bibnamefont{Torres}},
  \bibinfo{author}{\bibfnamefont{G.}~\bibnamefont{Molina-Terriza}},
  \bibnamefont{and} \bibinfo{author}{\bibfnamefont{L.}~\bibnamefont{Torner}},
  \bibinfo{journal}{J.\ Opt.\ B:\ Quantum\ Semiclass.\ Opt.}
  \textbf{\bibinfo{volume}{7}}, \bibinfo{pages}{235} (\bibinfo{year}{2005}).

\bibitem[{\citenamefont{Vaziri et~al.}(2003)\citenamefont{Vaziri, Pan,
  Jennewein, Weihs, and Zeilinger}}]{Vaziri03}
\bibinfo{author}{\bibfnamefont{A.}~\bibnamefont{Vaziri}},
  \bibinfo{author}{\bibfnamefont{J.-W.} \bibnamefont{Pan}},
  \bibinfo{author}{\bibfnamefont{T.}~\bibnamefont{Jennewein}},
  \bibinfo{author}{\bibfnamefont{G.}~\bibnamefont{Weihs}}, \bibnamefont{and}
  \bibinfo{author}{\bibfnamefont{A.}~\bibnamefont{Zeilinger}},
  \bibinfo{journal}{Phys.\ Rev.\ Lett.} \textbf{\bibinfo{volume}{91}},
  \bibinfo{pages}{227902} (\bibinfo{year}{2003}).

\bibitem[{\citenamefont{Collins et~al.}(2002)\citenamefont{Collins, Gisin,
  Linden, Massar, and Popescu}}]{Collins02}
\bibinfo{author}{\bibfnamefont{D.}~\bibnamefont{Collins}},
  \bibinfo{author}{\bibfnamefont{N.}~\bibnamefont{Gisin}},
  \bibinfo{author}{\bibfnamefont{N.}~\bibnamefont{Linden}},
  \bibinfo{author}{\bibfnamefont{S.}~\bibnamefont{Massar}}, \bibnamefont{and}
  \bibinfo{author}{\bibfnamefont{S.}~\bibnamefont{Popescu}},
  \bibinfo{journal}{Phys.\ Rev.\ Lett.} \textbf{\bibinfo{volume}{88}},
  \bibinfo{pages}{040404} (\bibinfo{year}{2002}).

\bibitem[{\citenamefont{Kaszlikowski et~al.}(2002)\citenamefont{Kaszlikowski,
  Kwek, Chen, \.Zukowski, and Oh}}]{Kaszlikowski02}
\bibinfo{author}{\bibfnamefont{D.}~\bibnamefont{Kaszlikowski}},
  \bibinfo{author}{\bibfnamefont{L.~C.} \bibnamefont{Kwek}},
  \bibinfo{author}{\bibfnamefont{J.-L.} \bibnamefont{Chen}},
  \bibinfo{author}{\bibfnamefont{M.}~\bibnamefont{\.Zukowski}},
  \bibnamefont{and} \bibinfo{author}{\bibfnamefont{C.~H.} \bibnamefont{Oh}},
  \bibinfo{journal}{Phys.\ Rev.\ A} \textbf{\bibinfo{volume}{65}},
  \bibinfo{pages}{032118} (\bibinfo{year}{2002}).

\bibitem[{\citenamefont{Clauser et~al.}(1969)\citenamefont{Clauser, Horne,
  Shimony, and Holt}}]{CHSH69}
\bibinfo{author}{\bibfnamefont{J.~F.} \bibnamefont{Clauser}},
  \bibinfo{author}{\bibfnamefont{M.~A.} \bibnamefont{Horne}},
  \bibinfo{author}{\bibfnamefont{A.}~\bibnamefont{Shimony}}, \bibnamefont{and}
  \bibinfo{author}{\bibfnamefont{R.~A.} \bibnamefont{Holt}},
  \bibinfo{journal}{Phys.\ Rev.\ Lett.} \textbf{\bibinfo{volume}{23}},
  \bibinfo{pages}{880} (\bibinfo{year}{1969}).

\bibitem[{\citenamefont{Ac\'{i}n et~al.}(2002)\citenamefont{Ac\'{i}n, Durt,
  Gisin, and Latorre}}]{Acin02}
\bibinfo{author}{\bibfnamefont{A.}~\bibnamefont{Ac\'{i}n}},
  \bibinfo{author}{\bibfnamefont{T.}~\bibnamefont{Durt}},
  \bibinfo{author}{\bibfnamefont{N.}~\bibnamefont{Gisin}}, \bibnamefont{and}
  \bibinfo{author}{\bibfnamefont{J.~I.} \bibnamefont{Latorre}},
  \bibinfo{journal}{Phys.\ Rev.\ A} \textbf{\bibinfo{volume}{65}},
  \bibinfo{pages}{052325} (\bibinfo{year}{2002}).

\bibitem[{\citenamefont{Vaziri et~al.}(2002{\natexlab{a}})\citenamefont{Vaziri,
  Weihs, and Zeilinger}}]{Vaziri02a}
\bibinfo{author}{\bibfnamefont{A.}~\bibnamefont{Vaziri}},
  \bibinfo{author}{\bibfnamefont{G.}~\bibnamefont{Weihs}}, \bibnamefont{and}
  \bibinfo{author}{\bibfnamefont{A.}~\bibnamefont{Zeilinger}},
  \bibinfo{journal}{Phys.\ Rev.\ Lett.} \textbf{\bibinfo{volume}{89}},
  \bibinfo{pages}{240401} (\bibinfo{year}{2002}{\natexlab{a}}).

\bibitem[{\citenamefont{Arlt et~al.}(1998)\citenamefont{Arlt, Dholakia, Allen,
  and Padgett}}]{Arlt98}
\bibinfo{author}{\bibfnamefont{J.}~\bibnamefont{Arlt}},
  \bibinfo{author}{\bibfnamefont{K.}~\bibnamefont{Dholakia}},
  \bibinfo{author}{\bibfnamefont{L.}~\bibnamefont{Allen}}, \bibnamefont{and}
  \bibinfo{author}{\bibfnamefont{M.~J.} \bibnamefont{Padgett}},
  \bibinfo{journal}{J.\ Mod.\ Opt.} \textbf{\bibinfo{volume}{45}},
  \bibinfo{pages}{1231 } (\bibinfo{year}{1998}).

\bibitem[{\citenamefont{Bazhenov et~al.}(1990)\citenamefont{Bazhenov,
  Vasnetsov, and Soskin}}]{Bazhenov90}
\bibinfo{author}{\bibfnamefont{V.~Y.} \bibnamefont{Bazhenov}},
  \bibinfo{author}{\bibfnamefont{M.~V.} \bibnamefont{Vasnetsov}},
  \bibnamefont{and} \bibinfo{author}{\bibfnamefont{M.~S.}
  \bibnamefont{Soskin}}, \bibinfo{journal}{JETP\ Lett.}
  \textbf{\bibinfo{volume}{52}}, \bibinfo{pages}{429} (\bibinfo{year}{1990}).

\bibitem[{\citenamefont{Vaziri et~al.}(2002{\natexlab{b}})\citenamefont{Vaziri,
  Weihs, and Zeilinger}}]{Vaziri02b}
\bibinfo{author}{\bibfnamefont{A.}~\bibnamefont{Vaziri}},
  \bibinfo{author}{\bibfnamefont{G.}~\bibnamefont{Weihs}}, \bibnamefont{and}
  \bibinfo{author}{\bibfnamefont{A.}~\bibnamefont{Zeilinger}},
  \bibinfo{journal}{J.\ Opt.\ B:\ Quantum\ Semiclass.\ Opt.}
  \textbf{\bibinfo{volume}{4}}, \bibinfo{pages}{S47}
  (\bibinfo{year}{2002}{\natexlab{b}}).

\bibitem[{\citenamefont{Gibson et~al.}(2004)\citenamefont{Gibson, Courtial,
  Padgett, Vasnetsov, Pas'ko, Barnett, and Franke-Arnold}}]{Gibson04}
\bibinfo{author}{\bibfnamefont{G.}~\bibnamefont{Gibson}},
  \bibinfo{author}{\bibfnamefont{J.}~\bibnamefont{Courtial}},
  \bibinfo{author}{\bibfnamefont{M.~J.} \bibnamefont{Padgett}},
  \bibinfo{author}{\bibfnamefont{M.}~\bibnamefont{Vasnetsov}},
  \bibinfo{author}{\bibfnamefont{V.}~\bibnamefont{Pas'ko}},
  \bibinfo{author}{\bibfnamefont{S.~M.} \bibnamefont{Barnett}},
  \bibnamefont{and}
  \bibinfo{author}{\bibfnamefont{S.}~\bibnamefont{Franke-Arnold}},
  \bibinfo{journal}{Optics\ Express} \textbf{\bibinfo{volume}{12}},
  \bibinfo{pages}{5448} (\bibinfo{year}{2004}).

\bibitem[{\citenamefont{Thew et~al.}(2004)\citenamefont{Thew, Ac\'{i}n,
  Zbinden, and Gisin}}]{Thew04}
\bibinfo{author}{\bibfnamefont{R.~T.} \bibnamefont{Thew}},
  \bibinfo{author}{\bibfnamefont{A.}~\bibnamefont{Ac\'{i}n}},
  \bibinfo{author}{\bibfnamefont{H.}~\bibnamefont{Zbinden}}, \bibnamefont{and}
  \bibinfo{author}{\bibfnamefont{N.}~\bibnamefont{Gisin}},
  \bibinfo{journal}{Phys.\ Rev.\ Lett.} \textbf{\bibinfo{volume}{93}},
  \bibinfo{pages}{010503} (\bibinfo{year}{2004}).

\bibitem[{\citenamefont{Chen et~al.}(2005)\citenamefont{Chen, Zhang, Bao,
  Schmiedmayer, and Pan}}]{Chen05}
\bibinfo{author}{\bibfnamefont{Z.-B.} \bibnamefont{Chen}},
  \bibinfo{author}{\bibfnamefont{Q.}~\bibnamefont{Zhang}},
  \bibinfo{author}{\bibfnamefont{X.-H.}~\bibnamefont{Bao}},
  \bibinfo{author}{\bibfnamefont{J.} \bibnamefont{Schmiedmayer}},
  \bibnamefont{and} \bibinfo{author}{\bibfnamefont{J.-W.} \bibnamefont{Pan}},
  \bibinfo{journal}{arXiv:quant-ph} p. \bibinfo{pages}{0501171}
  (\bibinfo{year}{2005}).

\end{thebibliography}
\end{document}